\documentclass[3p,twocolumn,times]{elsarticle}
\usepackage{amsmath}
\usepackage{amssymb}
\usepackage{amsthm}
\usepackage{graphics}
\usepackage{epsfig}
\usepackage[colorlinks=true,citecolor=blue,linkcolor=red,linktocpage=true,pagebackref=false]{hyperref}
\usepackage{soul} 
\usepackage[usenames, dvipsnames]{color}
\usepackage{braket}
\usepackage{changes}

\begin{document}

\begin{frontmatter}

\title{
Single-electron coherence: finite temperature versus pure dephasing
}

\author{Michael Moskalets}
\ead{michael.moskalets@gmail.com}
\address{Department of Metal and Semiconductor Physics, NTU ``Kharkiv Polytechnic Institute", 61002 Kharkiv, Ukraine}

\author{G\'eraldine Haack} 
\ead{geraldine.haack@gmail.com}
\address{Univ. Grenoble Alpes, INAC-SPSMS, F-38000 Grenoble, France}
\address{CEA, INAC-SPSMS, F-38000 Grenoble, France}

\date\today
\begin{abstract}
We analyze a coherent injection of single electrons on top of the Fermi sea in two situations, at finite-temperature and in presence of pure dephasing. Both finite-temperature and pure dephasing change the property of the injected quantum states from pure to mixed. However, we show that the temperature-induced mixedness does not alter the coherence properties of these single-electronic states. In particular two such mixed states exhibit perfect antibunching while colliding at an electronic wave splitter. This is in striking difference with the dephasing-induced mixedness which suppresses antibunching. On the contrary, a single-particle shot noise is suppressed at finite temperatures but is not affected by pure dephasing. This work therefore extends the investigation of the coherence properties of single-electronic states to the case of mixed states and clarifies the difference between different types of mixedness.
\end{abstract}

\begin{keyword}
Floquet scattering matrix \sep quantum transport \sep single-electron source \sep mixed state  \sep shot noise measurements
\PACS 73.23.-b \sep 72.10.-d \sep 73.63.-b
\end{keyword}

\end{frontmatter}


\section{Introduction}
\label{sec1}

Markus B\"{u}ttiker was a person who pioneered the branch of Physics presently known as Mesoscopics. Many of his predictions were successfully confirmed experimentally and gave birth to new exciting directions in Mesoscopics. One of such predictions directly relates to the topic of the present work.

In 1993 Markus B\"{u}ttiker, Harry Thomas, and Anna Pr\^{e}tre predicted that the low-frequency dissipative response of a single-channel mesoscopic capacitor is governed by the universal quantity, $R_{q} = h/(2e^{2})$, the charge relaxation resistance \cite{Buttiker:1993wh}. 
This value is for a single spin channel corresponding to half the value  of the von Klitzing resistance quantum, $R_{K} = h/e^{2}$ \cite{Klitzing:1980kw}. 
The charge relaxation resistance was suggested to be renamed {\it the B\"{u}ttiker resistance} \cite{Glattli:2014tu}, see also the comment in the work  by P. P. Hofer, D. Dasenbrook, and C. Flindt in this special issue. 

Besides being robust against interactions, see for instance Refs.~\cite{Nigg:2006kl,Mora:2010hw}, one of the key characteristic of this charge relaxation resistance is that it shows up if and only if the electrons conserve their quantum phase coherence while propagating through the sample \cite{Pretre:1996uw}. When the B\"uttiker resistance was confirmed experimentally in 2006 using a single-channel mesoscopic capacitor \cite{Gabelli:2006eg}, it then became clear that this setup could also serve as a coherent source of electrons. The year after, experimentalists made use of the relatively large energy level spacing in the capacitor to address a single quantum level. This ability allowed  them to operate the mesoscopic capacitor as a single-electron source, \textit{i.e.} a source which emits one particle at a time. 
With this experiment realized in 2007, Ref.~\cite{Feve:2007jx}, experimentalists achieved the emission of a periodic stream of alternating single electrons and single holes. \\
\indent
This experiment has then triggered a prolific both experimental 
\cite{Feve:2008gk,Mahe:2010cp,Hermelin:2011du,Parmentier:2012ed,Bocquillon:2012if,Bocquillon:2013ij,Bocquillon:2013dp,Fletcher:2013kt,Dubois:2013dv,Ubbelohde:2014eq,Jullien:2014ii,Freulon:2015jo,Waldie:2015vh,Vanevic:2015ue} 
and theoretical  
\cite{Moskalets:2008fz,Keeling:2008ft,Splettstoesser:2008gc,Splettstoesser:2009im,Sherkunov:2009jm,Sasaoka:2010ih,Albert:2010co,Lebedev:2011dh,Battista:2011jb,Albert:2011fx,  Haack:2011em, Grenier:2011js, Grenier:2011dv, Juergens:2011gu, Battista:2012db,Sherkunov:2012dg,Kashcheyevs:2012km,Battista:2013ew,Haack:2013ch,Grenier:2013gg,Lim:2013de,Ferraro:2013bt,Bocquillon:2013fp,Dasenbrook:2014db,Ludovico:2014de,Albert:2014bd,Gaury:2014jz,Thomas:2014jx,Xing:2014bo,Ferraro:2014ev,Gaury:2014ki,Moskalets:2014cg,Battista:2014di,Haack:2014ij,Hofer:2014jb,Rossello:2015iw,Thomas:2015jy,Ferraro:2015tg,Dasenbrook:2015ve,Tarasinski:2015wx,Thibierge:2015up,Dasenbrook:2015vd} 
activity, which constitutes now a fascinating and fast developing sub-field of Mesoscopics, which can be called {\it quantum coherent electronics}.  
Here quantum refers to the granular nature of charge and emphasizes the fact that the current through the conductor is generated by a flux of well separated  wave-packets.

Another step was then made theoretically by characterizing the coherence properties of the emitted electrons with the help of the first- and second order correlation functions \cite{Grenier:2011js, Haack:2013ch, Bocquillon:2013fp}, first introduced by Glauber in quantum optics. These works put forward the similar role played by the single-electron source in electronics compared to the already existing single-photon sources in quantum optics. This analogy also gave rise to an alternative denomination for this sub-field of Mesoscopics, electron quantum optics.

The past years have seen the realization of emblematic quantum optic experiments with single-electron sources. Analogs of the Hanbury--Brown and Twiss (HBT) experiment \cite{HanburyBrown:1956bi} and of the Hong-Ou-Mandel (HOM) experiment \cite{Hong:1987gm} with single electrons were respectively reported in Refs.~\cite{Bocquillon:2012if,Dubois:2013dv} and \cite{Bocquillon:2013dp,Dubois:2013dv}. 

Another important issue concerning the characterization of these single-electron sources concerns the nature of the quantum state which is emitted. This step was tackled in an ingenious experiment reported in Ref.~\cite{Jullien:2014ii}, where the wave function of a single electron emitted at low temperatures was measured. This result agrees well with the theoretical expectations from Refs.~\cite{Keeling:2006hq,Dubois:2013fs,Moskalets:2015vr}. 

The problem we address in the present work is the effect of a finite temperature onto the state emitted by a single-electron source. 
This problem is important at least for two reasons. 
First, experimentally, the electronic temperature is finite and without a clear understanding of the temperature effect, it is difficult to judge whether a given temperature can be considered as low or high. 
Second, the experiment from Ref.~\cite{Bocquillon:2012if} and the theory works \cite{Bocquillon:2012if,Dubois:2013fs} show that the shot noise caused by the scattering of single electrons at a quantum point contact gets suppressed with increasing electron temperature. 
This effect was explained as a result of antibunching of injected electrons with thermal excitations of the Fermi sea \cite{Bocquillon:2012if,Dubois:2013fs}.
As we  put forward in this work, there is an alternative interpretation of the shot noise reduction with a temperature increase. 

Our interpretation exploits the fact that the type of emitted quantum state (pure or mixed) is essential for characterizing the suppression or not of the shot noise. Our reasoning is based on the following arguments:
On one side, it is known that  energy relaxation processes suppress shot noise \cite{Blanter:2000wi}. 
On the other side, it is also known that when a pure state enters a region where it is subject to relaxation, then it becomes a mixed state. 
Therefore, a mixed state is expected to produce less shot noise compared to what is produced by a pure state.
Previous works have demonstrated that, in general, a pure state at zero temperature becomes mixed at finite temperatures, see for instance Ref.~\cite{Beenakker:2006vx}.

In the present work, we show that this statement is also valid for a quantum state emitted by single-electron sources. 
We also demonstrate that temperature-induced mixedness is not equivalent to pure-dephasing induced mixedness with respect to the coherence properties of the emitted quantum state. 
To this end, we compare an emitted quantum state in presence of finite-temperature and in presence of pure dephasing and emphasize striking differences in the shot noise suppression in an HBT and HOM experiments. Indeed, in the latter experiment, the shot noise relies on the phase coherence property between two incident single electrons emitted by different sources when they antibunch. As soon as these quantum states lose their phase coherence, in presence of pure dephasing, we observe a suppression of the antibunching. This has to be contrasted with the perfect   antibunching at finite-temperature. 
On the other hand, the phase coherence is irrelevant for the shot noise in HBT experiment. This is why the pure dephasing does not affect it. 
In contrast, the energy averaging required at finite temperatures suppresses the shot noise.
We formulate all our results in terms of the first-order correlation function of the emitted single electrons, which allows us to differentiate the effects caused by the suppression of phase coherence and averaging over energy.

The paper is organized as follows. 
In Sec.~\ref{sec2} we introduce the excess first-order correlation function $G ^{(1)}$, the basic quantity we use to characterize the state emitted by a single-electron source into an electronic waveguide. 
In the subsequent Sec.~\ref{sec3}, we relate $G ^{(1)}$ to measurable quantities such as the time-dependent current and the shot noise. 
The effect of a temperature of electrons in a waveguide onto $G ^{(1)}$ is discussed in detail in Sec.~\ref{sec4} and we compare these results with the pure dephasing situation in Sec.~\ref{sec5}.  
We conclude in Sec.~\ref{sec6}.

\section{Definition of $ G ^{(1)}$ }
\label{sec2}

The system we have in mind is a single-channel chiral waveguide of non-interacting and spinless electrons originating from a metallic contact. 
An electron system of a metallic contact is in equilibrium and is characterized by its Fermi distribution function $f$ with temperature $ \theta$ and chemical potential $ \mu$. 
As electronic waveguides one can use the edge states of conductors in the quantum Hall effect regime \cite{Klitzing:1980kw,Halperin:1982tb,Buttiker:1988fn} or of topological insulators \cite{Kane:2005hl,Bernevig:2006by,Buttiker:2009bg}.  
To inject electrons into a waveguide one can use a side-attached quantum dot, as experimentally realized in Refs.~\cite{Gabelli:2006eg,Feve:2007jx} in the  quantum Hall effect regime and theoretically suggested in Refs.~\cite{Hofer:2013cj,Inhofer:2013gd} for topological insulators. 
Another possibility is to use an in-line dynamic quantum dot \cite{Fletcher:2013kt,Waldie:2015vh,Leicht:2011ke}. A non-trivial way of generating of single-electron excitations was demonstrated in Ref.~\cite{Dubois:2013dv}, where, following a theoretical suggestion of Refs.~\cite{Levitov:1996ie,Ivanov:1997wz}, a periodic sequence of Lorentzian voltage pulses was applied directly to a metallic contact. A periodically working source emits a stream of particles. Here we assume that particles emitted during different periods do not overlap with each other.  

The first-order electronic correlation function \cite{Grenier:2011js, Haack:2013ch} in a waveguide after the source is defined in the full analogy with how it is done in optics \cite{Glauber:2006cy}, 

\begin{eqnarray}
{\cal G}^{(1)}\left(1;2  \right) =
\langle \hat\Psi^{\dag}(1)  \hat\Psi(2) \rangle ,  
\label{01}
\end{eqnarray}
\ \\ \noindent
where $\hat\Psi (j) \equiv \hat\Psi\left(x_{j},  t_{j} \right)$ is a single-particle electron field operator in second quantization evaluated at point $x_{j}$ and time  $t_{j}$  ($j=1,2$). 
The quantum-statistical average $\langle \dots \rangle$ is over the equilibrium state of electrons incoming from the metallic contact. 
The correlation function ${\cal G}^{(1)}$ contains information about electrons of the Fermi sea as well as about particles injected by the source. 

To access information solely about the particles emitted by the source let us introduce {\it the excess correlation function} \cite{Grenier:2011js,Grenier:2011dv,Haack:2013ch} evaluated as the difference of electronic correlation functions with the source on and off,

\begin{eqnarray}
G^{(1)}\left(1;2  \right) = {\cal G}^{(1)}_{on}\left(1;2  \right) - {\cal G}^{(1)}_{off} \left(1;2  \right) .
\label{excess}
\end{eqnarray}

The next step is to express $  G ^{(1)}$ in terms of some quantity characterizing an electronic source.  
To this end we first introduce the field operator in second quantization  $\hat\Psi\left(x_{j}, t_{j} \right)$ for electrons in an electrical conductor \cite{Buttiker:1992ge}. 
For chiral electrons it reads 

\begin{eqnarray}
\hat\Psi\left(x_{j}, t_{j} \right) = \int \frac{dE }{\sqrt{h v(E)} } e^{i\phi_{j}(E) } \hat b\left( E \right) .
\label{02}
\end{eqnarray}
\ \\ \noindent
Here $1/[h v(E)]$ is a one-dimensional density of states at energy $E$,  $\hat b(E)$ is an operator for electrons passed by the source, and the phase $\phi_{j}(E) = -E t_{j}/\hbar + k(E) x_{j}$.   
Then we use the Floquet scattering matrix $\hat S_{F}$ which characterizes  a source driven periodically by a potential with period ${\cal T} = 2 \pi/ \Omega$. It relates the $\hat b$-operators  introduced in Eq.~(\ref{02}) to the $\hat a$-operators which describe equilibrium electrons coming from the metallic contact \cite{Moskalets:2002hu,Moskalets:2011cw},

\begin{eqnarray}
\hat b(E) = \sum_{n} S_{F}\left( E, E_{n} \right) \hat a\left( E_{n} \right) . 
\label{03}
\end{eqnarray}
\ \\ \noindent 
Here $E_{n} = E + n \hbar \Omega$. 
The Floquet scattering matrix element $S_{F}\left( E_{n}, E \right)$ is a quantum mechanical amplitude for an electron with energy $E$ in a waveguide to absorb (or emit)  $n$ energy quanta $\hbar\Omega$ while passing through the source. 
If the source is off, then $S_{F}\left( E_{n}, E \right) = S(E) \delta_{n,0}$, where $S(E)$ is the stationary scattering amplitude. For a single-channel chiral case, it has an exponential form, $S(E) = e^{ i \varphi(E) }$. 

In order to perform a quantum-statistical average in Eq.~(\ref{01}) we use the following relation $\left\langle \hat a^{\dag}(E) \hat a(E^{\prime}) \right\rangle = f(E)\delta\left( E - E^{\prime} \right)$, which is valid since electrons in  a metallic contact are at equilibrium.
Finally, using the quantities introduced above, we represent the excess correlation function in terms of the Floquet scattering matrix of the source, 

\begin{eqnarray}
G^{(1)}(1;2) = 
\sum_{n,m=-\infty}^{\infty} 
\int \frac{dE }{ h v( E ) } f\left( E \right) 
e^{-i\phi_{1}(E_{n})} 
\nonumber \\
\label{05} \\
\times 
e^{i\phi_{2}(E_{m})}  
\left\{ 
S_{F}^{*}\left(E_{n},E  \right)
S_{F}\left(E_{m},E  \right)
- \delta_{m,0}\delta_{n,0} \right\} .
\nonumber 
\end{eqnarray}
\ \\ \noindent
In the case when the relevant energy scales of the problem (the amplitude of an applied voltage, the energy quantum $ \hbar \Omega$, etc.) all are small compared to the chemical potential$ \mu$, one can simplify the equation for $G ^{(1)}$.

\subsection{A linear dispersion approximation}

We employ the wide band approximation and treat the density of state as energy independent, $v(E) = v_{ \mu}$. 
We linearize the dispersion relation, 

\begin{eqnarray}
k\left( E_{n} \right) \approx k_{ \mu}  + \frac{ \epsilon + \hbar n\Omega }{ \hbar v_{ \mu} } ,
\label{06}
\end{eqnarray}
\ \\ \noindent
and represent the phase as follows,

\begin{eqnarray}
\phi_{j}(E_{n}) =   \phi_{ \mu,j} -  \bar t_{j} \frac{ \epsilon + \hbar n \Omega }{ \hbar } .
\label{07}
\end{eqnarray}
\ \\ \noindent
Here $ \phi_{ \mu,j} = - \mu t_{j}/\hbar + k_{ \mu} x_{j}$ and $k_{ \mu}$ are respectively  the phase factor and the wave vector for electrons at the energy  $E = \mu$, $ \epsilon = E - \mu$ is an energy counted from the chemical potential and $ \bar t_{j} = t_{j} - x_{j}/v_{ \mu}$ is an effective time. 
In addition it is convenient to introduce the scattering amplitude $S_{in}(t,E)$ \cite{Moskalets:2008ii}, such that its Fourier coefficients define the Floquet scattering amplitudes,  

\begin{eqnarray}
S_{F}(E_{n},E) = S_{in,n}(E) \equiv \int _{0}^{ {\cal T} } \frac{dt }{ {\cal T} } e^{in \Omega t} S_{in}(t,E) .
\label{08}
\end{eqnarray}
\ \\ \noindent
Then Eq.~(\ref{05}) can be cast into the following form, 

\begin{eqnarray}
G^{(1)}( \bar t_{1};\bar t_{2}) = \frac{ 1 }{hv_{ \mu} }
\int  d E    f\left( E \right) e^{i   \left( \bar t_{1} - \bar t_{2} \right) \frac{ E }{\hbar} } 
\nonumber \\
\label{09} \\
\times
\left\{  S_{in}^{*}( \bar t_{1}, E) S_{in}( \bar t_{2 }, E)    - 1 \right\} .
\nonumber 
\end{eqnarray}
\ \\ \noindent
This equation constitutes the basic equation we will use in the rest of the paper. 

The function $  G ^{(1)}$ encodes information about the state emitted by a single-electron source. 
To analyze this information one needs to relate $  G ^{(1)}$ to quantities  accessible experimentally. 
In the next section we present few examples of such relations.

\section{Applications of $  G ^{(1)}$}
\label{sec3}

\subsection{A time-dependent current}

\begin{figure}[t]
\centerline{
\includegraphics[width=80mm]{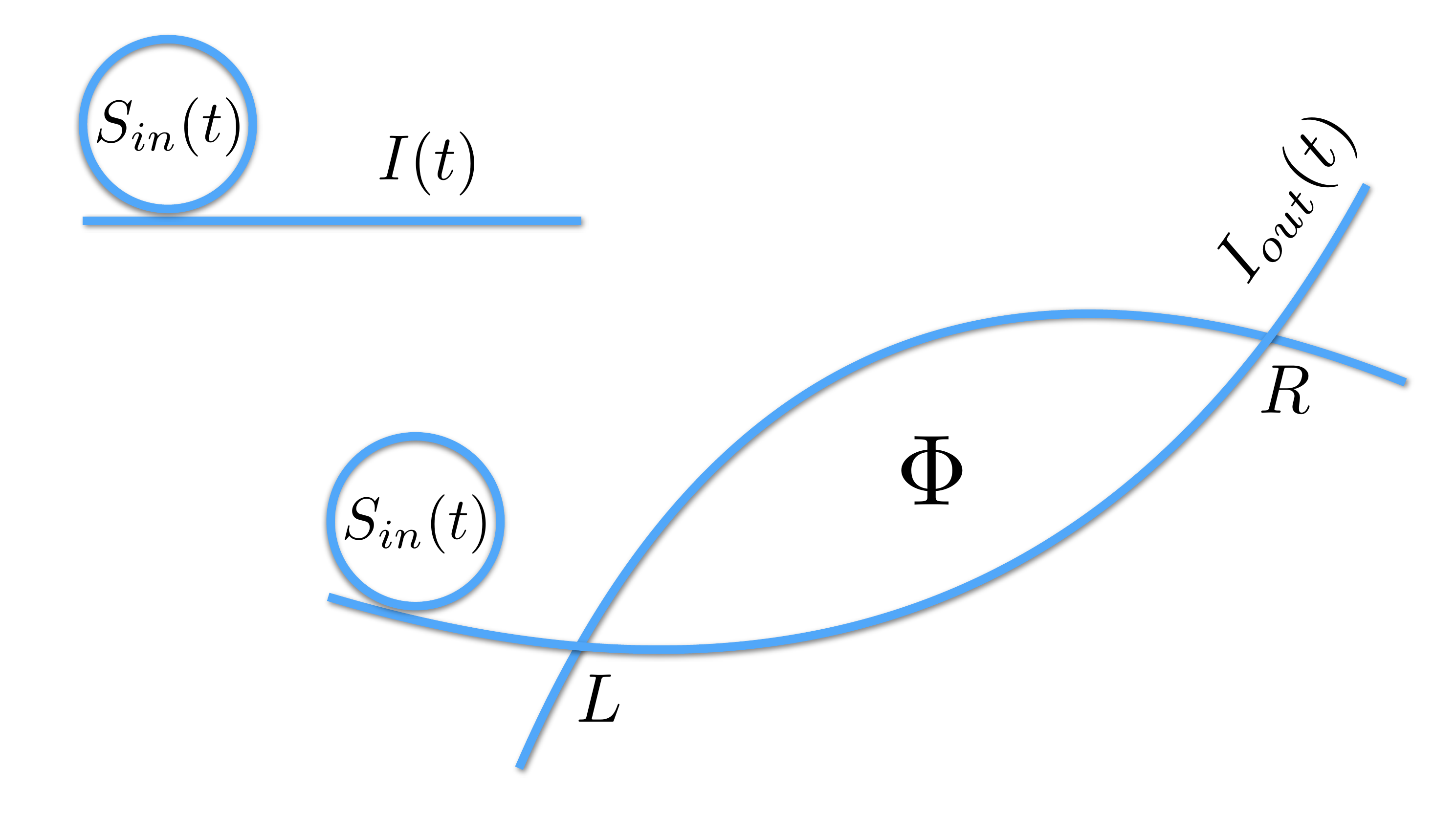}
}
\caption{(Color online) A single-electron source characterized by the scattering amplitude $S_{in}(t)$ injects periodically particles (electrons or electrons and holes) into a chiral wave-guide. {\bf Inset}: $I(t)$ is a current measured just behind the source. {\bf Main}: $I_{out}(t)$ is a current measured after an interferometer formed by two chiral wave-guides coupled at two quantum point contacts $L$ and $R$. $\Phi$ is a magnetic flux through the interferometer.}
\label{ses}
\end{figure}

A time-dependent current $I(t)$, measured just behind the source, see the inset to Fig.~\ref{ses}, is expressed in terms of the scattering amplitude \cite{Moskalets:2009dk},

\begin{eqnarray}
I(t) &=& \frac{e}{h}\int  dE f(E) 
\left\{  \left | S_{in}(t,E)  \right |^{2}  -1 \right\} .
\label{10}
\end{eqnarray}
\noindent \\
Comparing the equation above and Eq.~(\ref{09}) we see that $I(t)$ is given by the diagonal part, $ \bar t_{1} = \bar t_{2}\equiv \bar t$, of the correlation function, 

\begin{eqnarray}
I(t) = e v_{ \mu}  G ^{(1)}(t;t) .
\label{11}
\end{eqnarray}
\ \\ \noindent
Here $G^{(1)}\left( t;t \right)$ is evaluated at the same coordinates, $x_{1} = x_{2}$. 
This is why we use $t$ instead of $\bar t$. 

If the source emits a single particle with the wave function $\Psi(t)$, then 
$  G ^{(1)}(t;t) = \left | \Psi(t) \right |^{2}/v_{ \mu}$ \cite{Moskalets:2015vr}. 
Therefore, $I(t)$ provides information about a single-particle density, in particular about  the lifetime of the emitted quantum state.   

In order to access $G ^{(1)}(t_{1}; t_{2})$ at different times, $t_{1} \ne t_{2}$, it is necessary to measure an interference current, \textit{i.e.} to consider an interferometric setup.

\subsubsection{An interference current}

The current $I_{out}(t)$ after an electronic Mach-Zehnder interferometer \cite{Ji:2003ck}, see the main panel of Fig.~\ref{ses},  is expressed in terms of $G^{(1)}$ \cite{Haack:2013ch},

\begin{eqnarray}
I_{out}(t)  = R_{L}R_{R} \, I\left( t- \tau_{u} \right) +  
T_{L}T_{R} I\left( t- \tau_{d} \right)
\nonumber  \\
\label{12} \\
-  2 \gamma \, e v_{ \mu}\,\rm{Re}\left\{ e^{-i 2 \pi \frac{\Phi }{\Phi_{0} } } G^{(1)} (t- \tau_{u};t- \tau_{d}) \right\} . 
\nonumber 
\end{eqnarray} 
\noindent \\
Here $ \gamma = \sqrt{R_{L}R_{R}T_{L}T_{R}}$ with $R_{L/R}$ and $T_{L/R}= 1 - R_{L/R}$ are the reflection and transmission probabilities for the left $L$ and the right $R$ quantum point contacts of the interferometer. The traveling time along the upper/down arm of the interferometer is denoted $ \tau_{u/d}$, 
$\Phi$ is the magnetic flux through the interferometer and $\Phi_0=h/e$ is the magnetic flux quantum. 

The second line in Eq.~(\ref{12}) is an interference current. 
At $ \Phi$ multiple of the flux quantum, the interference current measured  as a function of $t$ and $ \Delta \tau$ gives access to the real part of $  G ^{(1)}$, while at $ \Phi - \Phi_{0}/4$ multiple of $\Phi_{0}$, the imaginary part of $  G ^{(1)}$ can be measured. 
The interference current, hence $G ^{(1)}(t_{1}; t_{2})$ at different times, provides information about the  single-particle coherence time \cite{Haack:2011em}, which depends on possibly present decoherence processes \cite{Haack:2013ch}.
 
Information  about the  two-particle coherence can be accessed via the measurement of a current correlation function in an electronic circuit with two sources. Let us here precise that a two-particle coherence has to be contrasted to a single-particle coherence. Whereas the latter one manifests itself in the interference current, the former one is due to the interference of two two-particle amplitudes and appears in the correlation noise in an HOM experiment. Two-particle amplitudes can interfere even when these two particles are not correlated.

\subsection{Shot noise}

The zero-frequency correlation function of currents outgoing to contacts $ \alpha$ and $ \beta$ of a multi-terminal electronic circuit fed by periodically working electron sources reads,

\begin{eqnarray}
{\cal P} _{\alpha \beta} &=& 
\frac{1 }{2 }
\int _{0} ^{ {\cal T}} \frac{dt }{ {\cal T} } \int_{ - \infty}^{ \infty} d \tau 
\nonumber \\
\label{p-noise} \\ 
&& \times
\left\langle \Delta \hat I _{\alpha}(t) \Delta \hat I _{\beta}(t - \tau) + \Delta \hat I _{\beta}(t - \tau)  \Delta \hat I _{\alpha}(t) \right\rangle.
\nonumber
\end{eqnarray}
Here $\Delta \hat I _{\alpha}(t)  = \hat I _{\alpha}(t)  - \left\langle \hat I _{\alpha}(t)  \right\rangle$ is the operator of current fluctuations; $\left\langle \dots \right\rangle$ stands for the quantum statistical average over the equilibrium state of electrons incoming from the contacts. 

We are interested in circuits whose output contacts are not directly connected to each other. 
In this case, the current cross-correlation function ${\cal P} _{\alpha \beta} $ with  $ \alpha \ne \beta$ is not sensitive to thermal fluctuations. 
It only depends  on the partitioning of incoming electron fluxes.  

If there is only one source which regularly injects particles into the circuit, the low-frequency current fluctuations are solely due to charge quantization and the corresponding noise is referred to as {\it the shot noise} \cite{Blanter:2000wi}. 
Let us briefly remind the reader that the shot noise arises because each electron can be transmitted to only one output contact. The current in this contact becomes therefore larger than its mean current, whereas the currents at the other output contacts become lower than their mean values. This effect leads to fluctuations in the current at each output contact.

If there is more than one input source, the partition noise can then be modified due to correlations between different simultaneously injected electrons. 
These correlations arise since the Pauli exclusion principle forbids two electrons in the same state to be scattered simultaneously to the same output contact. 

In the following we express ${\cal P} _{\alpha \beta}$ in terms of the excess first-order correlation function $G ^{(1)}$ of the injected electrons. 
This allows us first to show explicitly that, in the case of a single-electron source, the shot noise does not merely count the number of electrons emitted by the source but also depends on whether the emitted state is pure or mixed. 
Second, in the case of two sources connected to different input contacts, we clarify how the overlap of  single electrons at the wave-splitter affects the shot noise.

\subsubsection{One source}
\label{ss}

\begin{figure}[t]
\centerline{
\includegraphics[width=80mm]{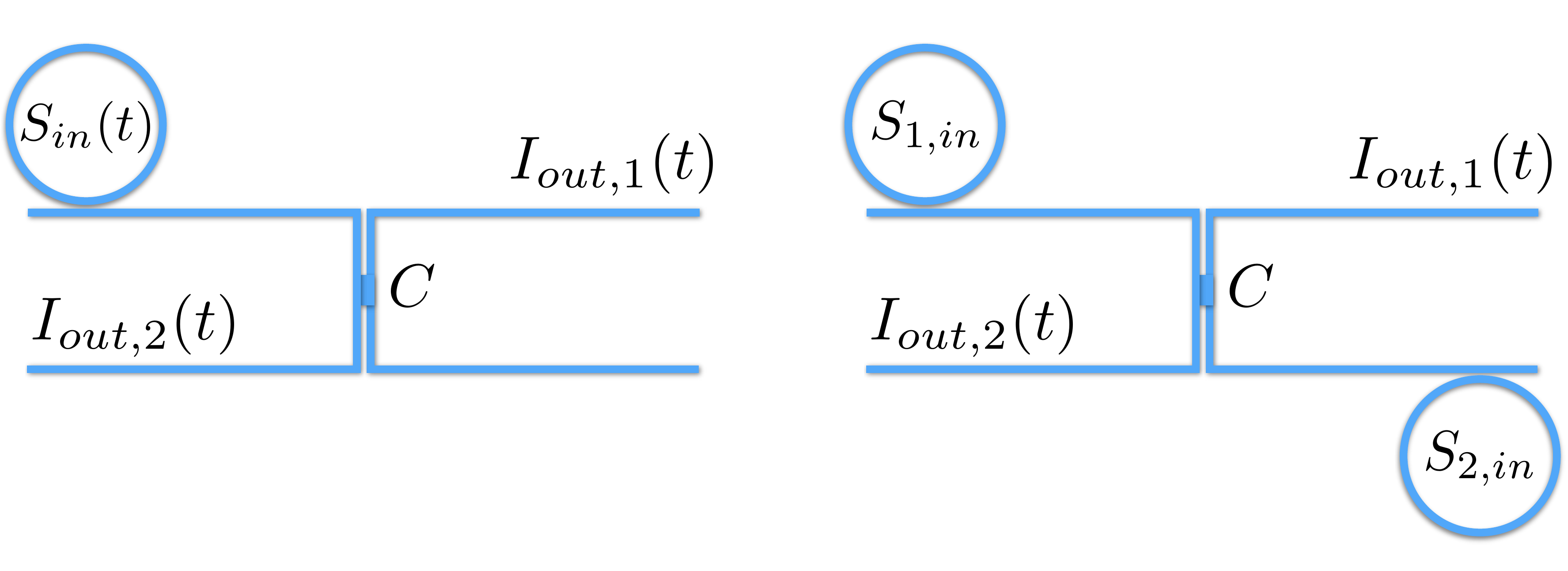}
}
\caption{(Color online) {\bf Left panel}: A wave-splitter setup. The periodic flux of particles injected by the source $S_{in}$ is split at the quantum point contact $C$. The time-dependent  currents measured at the corresponding outputs are denoted as $I_{out,1}$ and $I_{out,2}$, respectively. {\bf Right panel}: A collider setup. The particles injected by two sources $S_{1,in}$ and $S_{2,in}$ can collide at the quantum point contact $C$.}
\label{collider}
\end{figure}

Let us consider an electronic circuit, where the current injected by a source is splitted at a quantum point contact into two outgoing  wave-guides, see Fig.~\ref{collider}, left panel. 
Each input channel $j=1,2$ contains the Fermi sea with distribution function $f_{j}$. 
The zero-frequency correlation function of outgoing currents, ${\cal P} \equiv {\cal P}_{12}$, is expressed in terms of the scattering amplitude $S_{in}$ characterizing the source \cite{Moskalets:2011jx}, 

\begin{eqnarray}
\frac{  {\cal P} }{  {\cal P}_{0} } = 
- 2\int   \frac{dE }{ \hbar \Omega }\, 
\sum\limits_{q=- \infty }^\infty   
f_{ 1}\left( {E } \right)\left[f_{ 1}\left( {E } \right) - f_{ 2} \left( {E_{q} } \right) \right] 
\nonumber \\
\label{13} \\
\times
\int_{0}^{\cal T} \frac{dt}{\cal T} 
\int_{0}^{\cal T} \frac{dt^{\prime}}{\cal T} 
S^{*}_{in}\left( t,E \right) S_{in}\left( t^{\prime},E \right)  e^{ i q \Omega (t^{\prime} - t)}.
 \nonumber 
\end{eqnarray}
\ \\ \noindent
Here ${\cal P}_{0} = e^{2} R_{C}T_{C}/ {\cal T}$ is a zero-temperature shot noise caused by partitioning a single electron per period $ {\cal T}  = 2 \pi / \Omega$;   
$R_{C}$ and $T_{C} = 1 - R_{C}$ are the reflection and the transmission probabilities of the quantum point contact $C$, see Fig.~\ref{collider}. 
To simplify the notation, we omit a subscript for the outgoing channels. 

If two incoming channels have different temperatures, $f_{1} \ne f_{2}$, then the noise is not zero even if the source is off. 
In order to characterize the noise caused by a working source only, it is convenient to introduce the notion of {\it the excess noise}, which is defined as the difference of the shot noise with the source "on" and the one with the source "off". 
The excess current cross-correlation function reads 

\begin{eqnarray}
\frac{  {\cal P}^{ex} }{  {\cal P}_{0} } =  
2 \int   \frac{dE }{ \hbar \Omega }\, 
\sum\limits_{q=- \infty }^\infty    
f_{ 1}\left( {E } \right) f_{ 2} \left( {E_{q} } \right) 
\int_{0}^{\cal T} \frac{dt}{\cal T} 
\nonumber \\
\label{14} \\
\times
\int_{0}^{\cal T} \frac{dt^{\prime}}{\cal T} 
\left\{ S^{*}_{in}\left( t,E \right) S_{in}\left( t^{\prime},E \right)    - 1 \right\} e^{ i q \Omega (t^{\prime} - t)}.
\nonumber 
\end{eqnarray}
\ \\ \noindent
Note that deriving the equation above from Eq.~(\ref{13}), we have additionally used the unitarity of the scattering matrix,

\begin{eqnarray}
\int_{0}^{\cal T} \frac{dt}{\cal T} 
S^{*}_{in}\left( t,E \right) S_{in}\left( t, E_{p} \right) e^{- i p \Omega t}= \delta_{p,0} .
\label{16}
\end{eqnarray}
\ \\ \noindent
Using this equation with $p=0$, one can show that the term with $f_{1}^{2}$ in Eq.~(\ref{13}) does not contribute to Eq.~(\ref{14}).   

If the two incoming channels have the same temperatures, $f_{1} = f_{2} \equiv f$, then the excess current cross-correlation function ${\cal P}^{ex}$ can be expressed in terms of the excess single-particle correlation function $G ^{(1)}$.   
To show this we need, first, to consider an electronic collider, a circuit with two sources placed at two different input channels. 

\subsubsection{Two sources}

Let us consider an electron collider circuit with two sources, see Fig.~\ref{collider},  right panel.
Each source, $j=1,2$, is characterized by the scattering amplitude $S_{j,in}$. 
Now the excess current correlation function reads \cite{Moskalets:2011jx}, 

\begin{eqnarray}
\frac{  {\cal P}_{two}^{ex} }{  {\cal P}_{0} } &=& 
2 \int   \frac{dE }{ \hbar \Omega }\, 
\sum\limits_{q=- \infty }^\infty    
f_{ 1}\left( {E } \right)f_{ 2} \left( {E_{q} } \right)
\nonumber \\
\label{15} \\
&&\times
\iint_{0}^{\cal T} \frac{dt}{\cal T}  \frac{dt^{\prime}}{\cal T} 
\Bigg\{ 
S^{*}_{1,in}\left( t,E \right) S_{2,in}\left( t, E_{q} \right) 
 \nonumber \\
&&\times
S^{*}_{2,in}\left( t^{\prime},E_{q} \right) S_{1,in}\left( t^{\prime}, E \right) - 1 \Bigg\} e^{ i q \Omega (t^{\prime} - t)}.
\nonumber 
\end{eqnarray}
\ \\ \noindent
Here the subscript ``$two$'' shows that the two sources are present. 
Note that the time argument of the scattering amplitude takes into account the time of flight between the source and the quantum point contact $C$, see Fig.~\ref{collider}.

If the two sources are identical and they emit particles at the same time, $S_{1,in}(t,E) = S_{2,in}(t,E) \equiv S_{in}(t,E)$,  the excess noise vanishes,  ${\cal P}_{two}^{ex}  = 0$. This directly follows  from Eqs.~(\ref{15}) and (\ref{16}).

As a next step, we introduce {\it the correlation noise}, which is the difference of the excess noise caused by two working sources and the sum of excess noises generated by each source alone.  
The corresponding cross-correlation function is  
 
\begin{eqnarray}
\delta {\cal P} = {\cal P}_{two}^{ex} - {\cal P}_{1}^{ex} - {\cal P}_{2}^{ex} .
\label{17}
\end{eqnarray}
\ \\ \noindent
Here the subscript $j=1,2$ denotes which one of two sources is turned "on". 
Using Eqs.~(\ref{14}) and (\ref{15}), we find 
 
\begin{eqnarray}
\frac{ \delta {\cal P} }{  {\cal P}_{0} } =  
2 \int   \frac{dE }{ \hbar \Omega }\, 
\sum\limits_{q=- \infty }^\infty    
f_{ 1}\left( {E } \right)  f_{ 2} \left( {E_{q} } \right) 
\nonumber \\
\label{18} \\
\times
\iint_{0}^{\cal T} \frac{dt}{\cal T}   \frac{dt^{\prime}}{\cal T} 
\left\{ S^{*}_{1,in}\left( t,E \right) S_{1,in}\left( t^{\prime},E \right)    - 1 \right\}
\nonumber \\
\nonumber \\
\times
\left\{ S^{*}_{2,in}\left( t^{\prime},E_{q} \right) S_{2,in}\left( t,E_{q} \right) - 1 \right\} e^{ i q \Omega (t^{\prime} - t)}.
\nonumber 
\end{eqnarray}
\ \\ \noindent
Let us here emphasize that the correlation noise is a manifestation of a two-particle interference effect. As  nicely  demonstrated in the experiment \cite{Liu:1998wr}, the two-particle destructive interference effect is the manifestation of the Pauli exclusion principle which leads to the well-known antibunching of two electrons. We will make use of this argument below in Sec.~\ref{sec5} to emphasize the difference between temperature-induced mixedness and pure dephasing-induced mixedness.   

Equation (\ref{18})  can be expressed in terms of single-electron correlation functions for electrons emitted by the sources  when the period $ {\cal T}$ of each source is long enough such that the wave packets emitted by each  source during different periods do not overlap.

\subsubsection{A long period limit}
\label{lpl}

If the period $ {\cal T}  = 2 \pi / \Omega$ is long compared to all relevant time scales of the problem, that is,  the dwell time of the source, the lifetime of a wave packet, etc., then the energy quantum $ \hbar \Omega$ can be treated as  an  infinitesimal quantity and the sum over $q$ can be replaced by the integral over $ E_{q} = q \hbar \Omega$ according to the following rule:

\begin{eqnarray}
\sum\limits_{q=-\infty}^{\infty} \to \int  \frac{ d E_{q} }{ \hbar \Omega } .
\label{cont}
\end{eqnarray}
\ \\ \noindent 
Correspondingly, the integral over time effectively becomes running in infinite limits, $\int_{0}^{\cal T} dt \to \int_{- \infty}^{ \infty} dt$. 
In the following, we will not show explicitly these limits.
Equation~(\ref{18}) becomes 

\begin{eqnarray}
\frac{ \delta {\cal P} }{  {\cal P}_{0} } = \frac{2 }{h^{2} } 
\iint\limits dt dt^{\prime} \int   dE \int   dE_{q}
\nonumber \\
\label{20} \\
 \times
f_{ 1}\left( {E } \right) e^{ i (t - t^{\prime}) \frac{E }{ \hbar } }
\left\{ S^{*}_{1,in}\left( t,E \right) S_{1,in}\left( t^{\prime},E \right)    - 1 \right\}
\nonumber \\
\nonumber \\
 \times
f_{ 2} \left( {E_{q} } \right) e^{ i (t^{\prime} - t) \frac{E_{q}  }{ \hbar } }
\left\{ S^{*}_{2,in}\left( t^{\prime},E_{q} \right) S_{2,in}\left( t,E_{q} \right) - 1 \right\} .
\nonumber 
\end{eqnarray}
\ \\ \noindent
Comparing the equation above and Eq.~(\ref{09}), we arrive at the desired relation,

\begin{eqnarray}
\frac{ \delta {\cal P} }{  {\cal P}_{0} } = 2 v_{ \mu}^{2} \iint\limits dt dt^{\prime}  G ^{(1)}_{1}\left(t;t^{\prime} \right)  G ^{(1)}_{2}\left( t^{\prime};t  \right) ,
\label{21}
\end{eqnarray}
\ \\ \noindent
where $G ^{(1)}_{j}$ is the excess correlation function due to the source $j=1,2$. 
Since the correlation function satisfies the relation $  G ^{(1)}_{j}\left( t_{1}; t_{2} \right) = \left [  G ^{(1)}_{j}\left( t_{2}; t_{1} \right) \right]^{*}$, the above equation is real as it should be for a zero-frequency current correlation function, see Eq.~(\ref{p-noise}).

\subsubsection{Shot noise of a single source in terms of $G ^{(1)}$}

As announced at the end of Sec.~\ref{ss}, our goal is to relate the excess current cross-correlation function of a single source to the corresponding excess single-particle correlation function. To achieve it, we need both input channels having the same temperatures, $f_{1} = f_{2}$, and both sources being identical, $S_{1,in} = S_{2,in}$, hence  $  G ^{(1)}_{1} =  G ^{(1)}_{2} \equiv  G ^{(1)}$, see Eq.~(\ref{09}). 

In this case, as  already mentioned, the excess noise vanishes, ${\cal P}^{ex}_{two}=0$ and each source produces the same noise, ${\cal P}^{ex}_{1} = {\cal P}^{ex}_{2} \equiv {\cal P}^{ex}$. 
Therefore, the correlation noise is (minus) twice the noise due to a single source, hence $ \delta {\cal P} = - 2 {\cal P}^{ex}$, see Eq.~(\ref{17}). 
Now we can make use of Eq.~(\ref{21}) and express ${\cal P}^{ex}$ due to particles generated by a single source in terms of $  G ^{(1)}$:

\begin{eqnarray}
\frac{  {\cal P}^{ex} }{  {\cal P}_{0} } = - v_{ \mu}^{2} \iint\limits dt dt^{\prime} \left |   G ^{(1)}\left(t;t^{\prime} \right)  \right |^{2} . 
\label{22}
\end{eqnarray}
\ \\ \noindent
Taking into account the relation between the correlation function and the irreducible part of the two-particle distribution function \cite{Moskalets:2015vr}, we conclude that Eq.~(\ref{22}) agrees with the result of Ref.~\cite{Moskalets:2006gd} on the noise produced by adiabatic quantum pumps at zero temperature. 

We stress that Eq.~(\ref{22}) was derived under the condition that both input channels have the same electronic temperature. 
Therefore, the corresponding shot noise is solely due to injected particles. 
There is no effect caused by the antibunching of injected particles with thermal excitations of the Fermi sea coming from the other input channel.  
Indeed, all thermal excitations coming from one input channel do experience antibunching with the thermal excitations coming from the other input channel. One particle cannot antibunch with two particles at once \cite{Moskalets:2011jx}.

\subsubsection{Shot noise of two sources}

If the two sources are different, $  G ^{(1)}_{1} \ne  G ^{(1)}_{2}$, but the temperatures of both input channels are the same,  we can use Eqs.~(\ref{22}) and (\ref{21}) in (\ref{17}) and obtain,

\begin{equation}
\frac{  {\cal P}^{ex}_{two} }{  {\cal P}_{0} } = - v_{ \mu}^{2} \iint\limits dt dt^{\prime} \left |   G ^{(1)}_{1}\left(t;t^{\prime} \right) - G ^{(1)}_{2}\left(t;t^{\prime} \right)  \right |^{2} .
\label{23}
\end{equation}
\ \\ \noindent
The negative sign of the fermionic cross-correlation function is expected from a non-interacting theory in the absence of relaxation \cite{Buttiker:1992ge,Texier:2000vh}. 

Equation (\ref{23}) is manifestly zero if the states emitted by the two electronic   sources are identical and if they reach the beam splitter at the same time: $G ^{(1)}_{1} = G ^{(1)}_{2}$. 
This means that the two emitted states demonstrate a perfect antibunching  while scattering at the wave-splitter, that results in vanishing the shot noise \cite{Liu:1998wr,Henny:1999tb,Oliver:1999ws,Oberholzer:2000wx}.
This is a manifestation of an electronic analogue \cite{Bocquillon:2013dp,Dubois:2013dv,Olkhovskaya:2008en,Jonckheere:2012cu,Ferraro:2014if,Iyoda:2014cf,Khan:2014bz,Ferraro:2014ux} of the Hong-Ou-Mandel effect \cite{Hong:1987gm}, well known in quantum optics. 

In the particular case when each source emits a pure single-particle state, \textit{i.e.} $  G ^{(1)}_{j}\left(t;t^{\prime} \right) = \Psi_{j}^{*}(t) \Psi_{j}\left( t^{\prime} \right)/ v_{ \mu}$ with $\Psi_{j}$ being the wave function of a single-electron excitation emitted by the source $j = 1,2$, the excess current cross-correlation function can be expressed in terms of the overlap integral $J = \int dt \Psi_{1}^{*}(t) \Psi_{2}(t)$. We use a normalization $\int dt \left | \Psi_{j}(t) \right |^{2} = 1$ and obtain from Eq.~(\ref{23}) the desired relation, 

\begin{eqnarray}
\frac{ {\cal P}^{ex}_{two} }{  {\cal P}_{0} }  = -2\left( 1 -  \left | J \right |^{2}\right), 
\label{23-1}
\end{eqnarray}
\ \\ \noindent
see for example Refs.~\cite{Bocquillon:2013fp,Dubois:2013fs,Moskalets:2013dl}. 
In general, we do not expect such a relation in the case of a mixed state, in particular, at finite temperatures. 
However, in the case of colliding levitons the temperature effect is decoupled from the overlap effect \cite{Dubois:2013fs}.

\subsection{Shot noise of a single source and the number of particles}

Equation (\ref{22}) relates the shot noise to the correlation function. 
In the case of a pure state the shot noise can be related to the rate at which the particles are emitted. 

\subsubsection{Pure state}

If the state emitted by the source is pure, then  Eq.~(\ref{22}) can be further simplified. 
For this, we note that the single-particle correlation function for a pure state satisfies the following identity

\begin{eqnarray}
v_{ \mu} \int dt  G ^{(1)}\left( t_{1}; t \right)  G ^{(1)}\left( t; t_{2} \right) =  \pm G ^{(1)}\left( t_{1}; t_{2} \right), 
\label{24}
\end{eqnarray}
\ \\ \noindent
where the sign $+$ stays for an electronic state and the sign $-$ is for a hole state.
Inserting the above equation into Eq.~(\ref{22}), we obtain

\begin{eqnarray}
\frac{  {\cal P}^{ex} }{   \left( -  {\cal P}_{0} \right)  } = N\,.
\label{25}
\end{eqnarray}
\ \\ \noindent
Here we introduce the (statistical mean) number of particles $N$ emitted during one  period:

\begin{eqnarray}
N = \pm v_{ \mu} \int_{0}^{\cal T} dt  G ^{(1)}\left( t;t \right) = \pm \frac{1 }{e } \int_{0}^{\cal T} dt I(t),
\label{26}
\end{eqnarray}
\ \\ \noindent 
and we have restored the proper limits of the time integral for clarity.  
The symbol $e$ stays for an electron charge and the sign $+$/$-$ is  for an electron/ a hole state.  
If a pure emitted state contains both electrons and holes, then $N$ is the sum of the number of electrons and holes emitted during one period. 

On the contrary, if the emitted state is  mixed, Eq.~(\ref{24}) does not hold anymore and  the original equation (\ref{22}) should be used instead of Eq.~(\ref{25}). 
In this case, there is no a direct relation between the number of emitted particles and the shot noise.

\subsubsection{Mixed state} 
\label{snms}

To clarify how the shot noise is related to the particle number in the case of a mixed state, we discuss few examples.
Let us start from a simple circuit, where a mixed state can be created: a wave splitter with transmission probability $T$. 
A single-particle state is in one input channel of the wave splitter and the vacuum state is in the other one.  
Each of the output states is a mixture of a single-particle state and the vacuum state. 
Let us take an output channel, where a single-particle state appears with probability $T$. 
If such a mixed state is directed to the second wave splitter, then the current cross-correlation function is reduced by the factor $T^{2}$ compared to the case where a single particle state is directly partitioned at the second wave splitter \cite{Oberholzer:2000wx,Texier:2000vh,Moskalets:2011jx}. 
In this case, the rate at which the particles impinge on the second wave splitter is reduced by a factor $T$ while the shot noise is reduced by a factor $T^{2}$.  

If the mixed state is created without changing the arrival rate of the electrons  at the wave splitter, the shot noise is changed or not (compared to the  case of a pure state) depending on the nature of decoherence processes. 
A pure dephasing does not affect the shot noise while inelastic scattering suppresses the shot noise \cite{Blanter:2000wi}. 

A finite temperature in itself is not considered as a decoherence factor but it also turns a pure state into a mixed state  \cite{Beenakker:2006vx}. 
As we will show in the next section,  a temperature-induced mixedness suppresses the shot noise (without any suppression in the particle number). To emphasize the difference between a temperature effect and decoherence we also present calculations for a pure dephasing model.

\section{$G ^{(1)}$ at finite temperatures} 
\label{sec4}

To highlight the effect of a non-zero temperature, we focus from now on a particular model for a single-electron source. 
Namely, we consider the source of levitons from Refs.~\cite{Dubois:2013dv,Jullien:2014ii}, where  the stream of single electrons  is excited out of a metallic contact with the help of a periodically repeated Lorentzian voltage pulse \cite{Levitov:1996ie,Ivanov:1997wz}. 
If the half-width of a leviton $  \Gamma _{\tau}$ is much shorter compared to  the period, $\Gamma _{\tau} \ll {\cal T}$,  electrons excited during different periods do not overlap and they are statistically independent of each other. 
In this case we can use an approximation introduced in Sec.~\ref{lpl} and restrict ourselves to only one period, which is formally treated as an infinite interval. 

Explicitly, the Lorentzian voltage pulse applied to the metallic contact to create a single-electron excitation has the form: $eV_{V_{L}}(t) = 2 \hbar  \Gamma _{\tau}/\left( t^{2} +  \Gamma _{\tau}^{2} \right)$.    
We denote the corresponding energy-independent scattering amplitude  and the excess correlation function as $S_{in} \equiv S_{V_{L}}$ and $  G ^{(1)}_{V_{L} }$ respectively. 
Up to some irrelevant phase factor, the scattering amplitude reads 

\begin{eqnarray}
S_{V_{L}}(t) = e^{ - i \frac{e }{ \hbar} \int _{}^{t } dt^{\prime} V_{L}(t^{\prime}) } = \frac{ t + i  \Gamma _{\tau} }{ t - i  \Gamma _{\tau} } .
\label{27}
\end{eqnarray}
\ \\ \noindent
The maximum of a single-particle density is at $t=0$. 

To cast Eq.~(\ref{09}) into a form convenient to the subsequent discussion, we use a substitution $ x e^{xE} \to d \left( e^{xE} \right)/dE$ (with $x=i\left( \tau_{1} - \tau_{2} \right)/ \hbar$) and integrate by parts over $E$. 
Then we obtain, 

\begin{eqnarray}
G^{(1)}_{V_{L} }( t_{1};t_{2}) &=& 
\int  d E \left( - \frac{   \partial f }{  \partial E } \right) G^{(1)}_{ V_{L},E}( t_{1};t_{2}) ,
\nonumber \\
\label{28} \\
G^{(1)}_{ V_{L},E}( t_{1};t_{2}) &=&
\frac{1  }{  v_{ \mu} } \Psi_{V_{L},E}^{*}\left( t_{1} \right) \Psi_{V_{L},E}\left( t_{2} \right) ,
\nonumber 
\end{eqnarray}
\ \\ \noindent
where $\Psi_{V_{L},E}(t) = e^{- i  t\frac{ E }{\hbar} }\sqrt{  \Gamma _{\tau}/ \pi} / \left( t - i  \Gamma _{\tau} \right)$ is the wave function of a leviton \cite{Keeling:2006hq}.  
Since $G^{(1)}_{ V_{L},E}$ satisfies Eq.~(\ref{24}), it can be interpreted as a correlation function of a leviton being in a pure state with the filling frequency $E/ \hbar$ and the envelop function $\sqrt{  \Gamma _{\tau}/ \pi} / \left( t - i  \Gamma _{\tau} \right)$.

\subsection{Temperature-induced mixedness}

At zero temperature $\partial f / \partial E = - \delta\left( E - \mu \right) $, where $ \delta(x)$ is the Dirac delta function. 
Therefore $G^{(1)}_{V_{L}} = G^{(1)}_{V_{L},\mu}$ and the state emitted by the source is a pure state with the filling frequency $ \mu/ \hbar$. 

On the contrary, at non-zero temperatures, the correlation function $  G ^{(1)}_{V_{L}}$, Eq.~(\ref{28}), does not satisfy Eq.~(\ref{24}). 
Therefore, the emitted state is a mixed state with the probability density of appearance of a pure state $\Psi_{L,E}$ being 

\begin{eqnarray}
p_{ \theta}(E) = -  \partial f (E) /  \partial E .
\label{29}
\end{eqnarray}
\ \\ \noindent
Notice that the different components of this mixed state are not orthogonal to each other, 

\begin{eqnarray}
\int dt \Psi_{V_{L},E}^{*} \Psi_{V_{L},E^{\prime}} = e^{- \frac{\left | E - E^{\prime} \right | }{ 2 {\cal E}_{L} } } .  
\label{30}
\end{eqnarray}
\ \\ \noindent
where $ {\cal E}_{L} = \hbar /(2  \Gamma _{\tau})$ is the mean energy of a leviton \cite{Keeling:2006hq}.
However, if the difference of the filling frequencies of two states exceeds $1/  \Gamma _{\tau}$, then these states are effectively orthogonal to each other.  

To differentiate pure and mixed states of a leviton, let us introduce a time-dependent purity coefficient which we define as follows:

\begin{eqnarray}
\mathrm{P_{V_{L}}}\left( t_{1}, t_{2} \right) &=& 
\frac{ v_{ \mu} \int dt  G ^{(1)}_{V_{L}}\left( t_{1}; t \right)  G ^{(1)}_{V_{L}}\left( t; t_{2} \right) }{ G ^{(1)}_{V_{L}}\left( t_{1}; t_{2} \right) }
\nonumber \\
\label{31} \\
&=& \frac{ \int  dt  \frac{  \Gamma _{\tau}/ \pi }{ t^{2} +  \Gamma _{\tau}^{2} } \eta\left( \frac{  t_{1} - t }{ \tau_{ \theta} }  \right)  \eta\left( \frac{  t - t_{2} }{ \tau_{ \theta} }  \right)  }{ \eta\left( \frac{  t_{1} - t_{2} }{ \tau_{ \theta} }  \right) } .
\nonumber 
\end{eqnarray}
\ \\ \noindent
Here $\eta(x) = x/ \sinh(x) \equiv \int dE (-  \partial f/  \partial E) \exp\left( i x \tau_{ \theta} \frac{E  }{ \hbar } \right)$ and the thermal time $ \tau_{ \theta} = \hbar / ( \pi k_{B} \theta )$, where $k_{B}$ is the Boltzmann constant.

\begin{figure}[t]
\includegraphics[width=75mm]{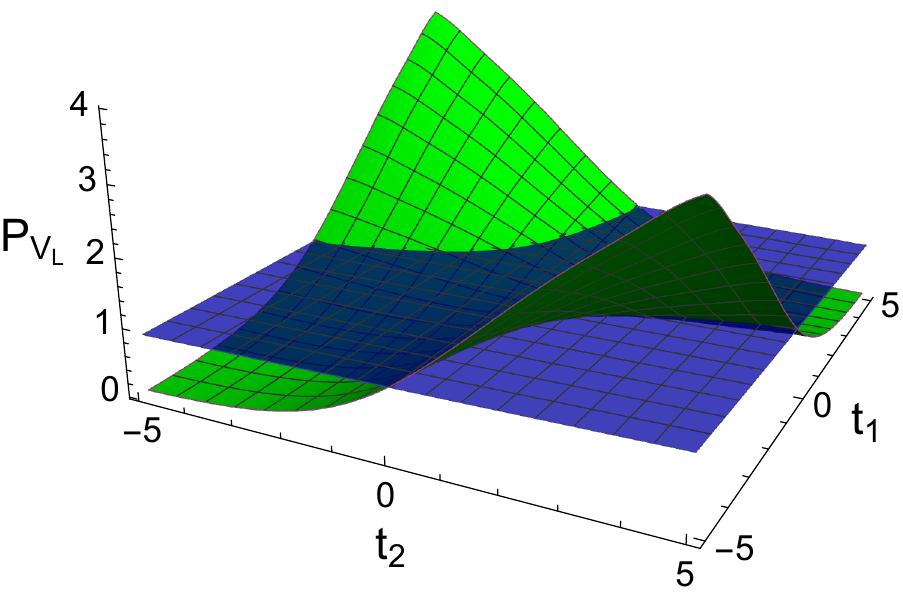}
\caption{(Color online) The purity coefficient $\mathrm{P_{V_{L}}}$ for levitons emitted into the Fermi sea with a finite temperature is shown as a function of  two times $t_{1}$ and $t_{2}$ normalized to the leviton's half-width $\Gamma_{ \tau}$, see Eq.~(\ref{30}). The blue transparent surface calculated for a small temperature, $\tau_{ \theta}/\Gamma _{\tau}=10$, demonstrates that the state of a leviton is pure since the purity coefficient is close to one. While the green opaque surface calculated for a relatively large temperature, $\tau_{ \theta}/\Gamma _{\tau}=1$, demonstrates that the state is mixed since the purity coefficient deviates strongly from one. The maximum of leviton's density is at $t_{1} = t_{2} = 0$.}
\label{purity}
\end{figure}

The purity coefficient is shown in Fig.~\ref{purity}.  
For a pure sate, the coefficient $\mathrm{P_{V_{L}}}\left( t_{1}, t_{2} \right) = 1$ for any times $t_{1}$ and $t_{2}$, while for a mixed state, it differs from $1$ for some $t_{1}$ and $t_{2}$. 
Figure  \ref{purity} shows that the state of a leviton is pure if the thermal time exceeds its duration, $ \tau_{ \theta} \gg  2\Gamma _{\tau}$ (see the blue transparent surface for $\tau_{ \theta} = 10  \Gamma _{\tau}$).   
On the contrary, at higher temperatures when  $ \tau_{ \theta} \le  2\Gamma _{\tau}$,  the state of a leviton becomes mixed (see the green opaque surface for $\tau_{ \theta} =  \Gamma _{\tau}$).

With this, we could answer the first main question we formulated in the introduction: how to consider an electronic temperature as high or low? The answer lies in the comparison between the thermal time $\tau_\theta$ and the life time of the single-particle state $\Gamma_\tau$ as discussed above.

\subsection{Temperature suppression of the shot noise}

Inserting Eq.~(\ref{28}) into Eq.~(\ref{22}), the excess current cross-correlation function in the case of a periodic stream of levitons reads:

\begin{eqnarray}
\frac{  {\cal P}^{ex}_{L} }{ \left( - {\cal P}_{0}  \right) } =  
 \iint dE dE^{\prime} p_{ \theta}(E) p_{ \theta}(E^{\prime}) e^{- \frac{\left | E - E^{\prime} \right | }{ {\cal E}_{L} } } .
\label{32}  
\end{eqnarray}
\noindent
Note that the factor $\exp\left(- \left | E - E^{\prime} \right | / {\cal E}_{L}  \right)$ arises due to the  non-orthogonality of the different components of a mixed state, see Eq.~(\ref{30}).

\begin{figure}[t]
\includegraphics[width=75mm]{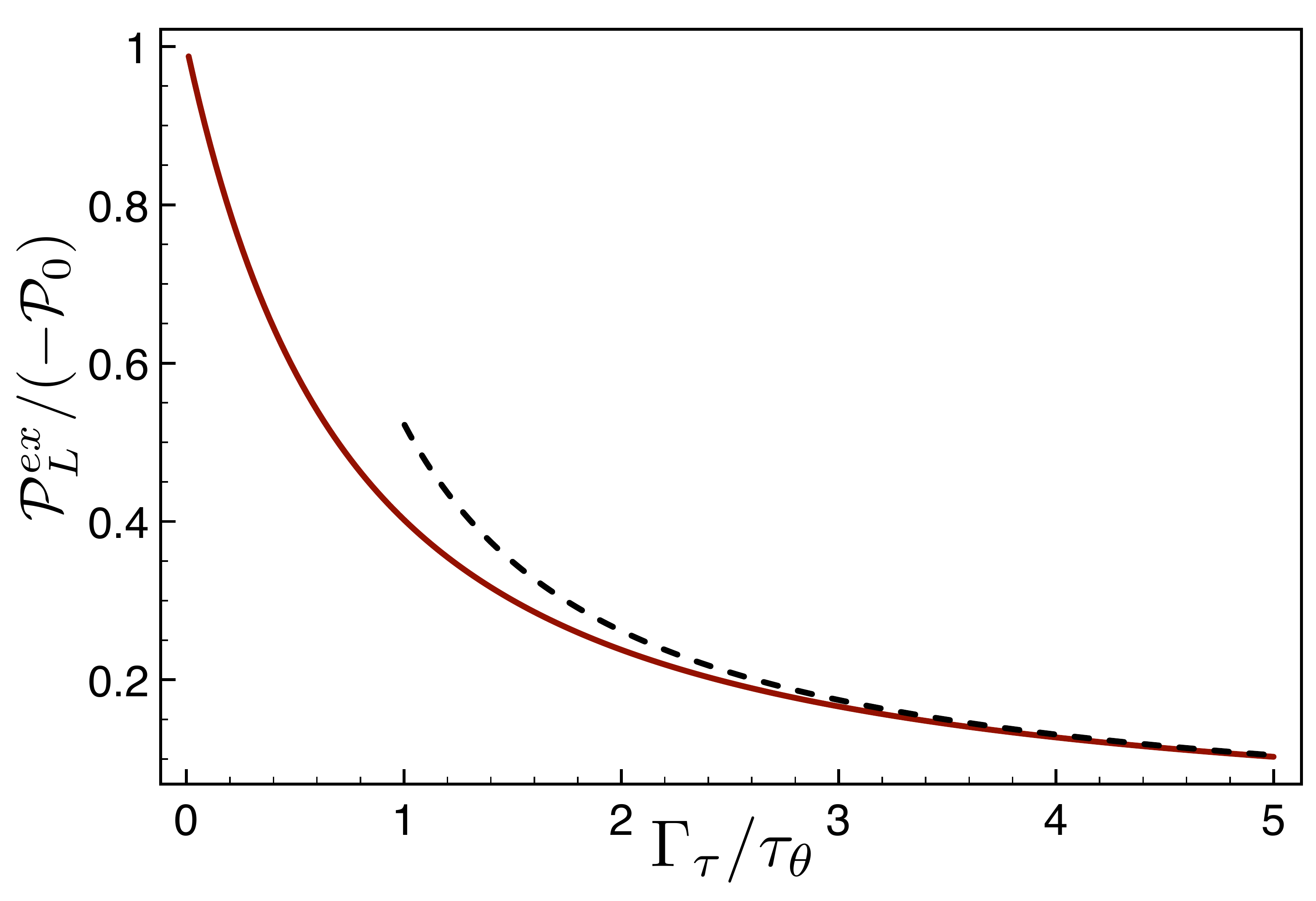}
\caption{(Color online) The excess current cross-correlation function, $ {\cal P}^{ex}_{L}/(-  {\cal P}_{0})$,  Eq.~(\ref{32}), is shown (a solid red line) as a function of the ratio of the half-width of a leviton $  \Gamma _{\tau}$ and the thermal time $\tau_{ \theta} = \hbar/( \pi k_{B} \theta)$. The asymptotics given in Eq.~(\ref{33}) is shown as a dashed black line.}
\label{noise}
\end{figure}

The function ${\cal P}^{ex}_{L} $ is shown in Fig.~\ref{noise}. 
At low temperatures, when $ \tau_{ \theta} \gg  \Gamma _{\tau}$, as we already mentioned, the leviton is emitted in a pure state and  ${\cal P}^{ex}_{L} / \left( -  {\cal P}_{0} \right) = 1$. 
While with increasing temperature the current cross-correlation function  decreases. 
At relatively high temperatures, when $ \tau_{ \theta} \le  \Gamma _{\tau}$, it  can be represented as follows,

\begin{eqnarray}
\left. \frac{  {\cal P}^{ex}_{L} }{  \left( -{\cal P}_{0}  \right)}  \right |_{k_{B} \theta \gg  \hbar/  \Gamma _{\tau} } \approx 
2 {\cal E}_{V_{L}}  \int dE p_{ \theta}^{2}(E) = \frac{ \pi }{ 6 } \frac{\tau_{ \theta} }{  \Gamma _{\tau}  } .
\label{33}
\end{eqnarray}
\ \\ \noindent
The corresponding dependence is shown in Fig.~\ref{noise} by the dashed line. 
Clearly it is a good approximation to the exact result, Eq.~(\ref{32}), already at $  \Gamma _{\tau}/ \tau_{ \theta} \ge 3$. 

Equation (\ref{33}) can be interpreted in the following way. 
At high temperatures the states with energy from a wide interval (of the order of few $k_{B} \theta$ near the chemical potential $ \mu$) do contribute to the state of an emitted particle, see Eq.~(\ref{28}). 
The two states are correlated if their energies do not differ by more than $ 2 {\cal E}_{L}$, see Eq.~(\ref{30}). 
This property allows us to replace the double-energy integral in Eq.~(\ref{32}) by a  single-energy integral in Eq.~(\ref{33}) with the extra factor $2 {\cal E}_{V_{L}}$  being the correlation energy for the states $\Psi_{L,E}$s. 
Therefore, at high temperatures, the state of a leviton can be viewed as a mixture of effectively uncorrelated pure states arising with probability $\bar p_{ \theta}(E) = 2 {\cal E}_{V_{L}} \left( -  \partial f (E) /  \partial E \right)$. 
Each of these states occupies an effective volume $2 {\cal E}_{V_{L}}$ in the energy space. 

This picture, which emerges as a result of energy averaging, allows us to express measurable quantities for a leviton  in a  mixed state via the corresponding quantities calculated for a pure state. 
For example, Eq.~(\ref{28}) becomes $  G ^{(1)}_{V_{L}} = \int dE/(2 {\cal E}_{V_{L}})\bar p_{ \theta}(E)  G ^{(1)}_{L,E}$, where $G ^{(1)}_{L,E}$ is for a    pure state with energy $E$.  
While Eq.~(\ref{33}) can be rewritten as ${\cal P}^{ex}_{L}   = \int dE/(2 {\cal E}_{V_{L}}) \bar p_{ \theta}^{2}(E)  \left( -{\cal P}_{0}  \right)$, with $\left( -{\cal P}_{0}  \right)$ being the excess current correlation function calculated for a leviton in a pure state. 
Therefore, the component of a mixed state appearing with probability $\bar p_{ \theta}$ contributes to noise with an amount proportional to $\bar p_{ \theta}^{2}$. 
Though at a finite temperature the number of emitted particles is not changed,  $\int dE/(2 {\cal E}_{V_{L}}) \bar p_{ \theta}(E) = 1$, the noise is reduced since $\int dE/(2 {\cal E}_{V_{L}}) \bar p_{ \theta}^{2}(E) <  1$.

\subsection{Antibunching of electrons emitted at finite temperatures}

The two identical single-electron states $\left(G ^{(1)}_{1} =  G ^{(1)}_{2}\right) $ demonstrate perfect antibunching even if they are emitted at finite temperatures, that is, even if they are mixed states. 
As a result of antibunching the shot noise is completely suppressed, ${\cal P}^{ex}_{two}=0$, see Eq.~ (\ref{23}). 
This was observed experimentally in Ref.~\cite{Dubois:2013dv} in HOM-type measurements of levitons. 
The corresponding prediction was made in Ref.~\cite{Dubois:2013fs}, where it was also shown theoretically that the HOM noise at zero-time delay vanishes at all temperatures for sine wave voltage pulses. 
Our equation (\ref{23}) generalizes this result to arbitrary voltage pulses and to different types of coherent single-electron sources.

The vanishing of the shot noise means that a positive correlation noise, see Eq.~(\ref{21}), completely compensates negative single-particle contributions, see Eq.~(\ref{22}). 
As we showed above, with increasing temperature, the single-particle contributions get suppressed. 
Therefore, the correlation noise also gets suppressed in order to maintain a zero total shot noise. 
As we already mentioned, the correlation noise originates from a two-particle interference. 
Therefore, one can suppose that a two-particle interference is suppressed at finite temperatures, i.e., that a finite temperature is a decoherence factor.   
However this is not the case, since a two-particle interference contribution to noise (the correlation noise) is suppressed in a similar way as a single-particle contribution, which does not rely on any interference effect. 
This fact indicates that a temperature-induced mixedness does not destroy the coherence properties of the single-electron state. 
What is suppressed is not an interference ability but the correlation noise, which is given by the sum over the components of a mixed state.  

Note that the exact compensation of both a correlation contribution and a single-particle contribution to the shot noise is not specific to the source of levitons: it is valid for any coherent electron source at zero as well as finite temperatures.

\subsection{Second-order correlation function $G ^{(2)}$ for electrons emitted at finite temperatures}

The second-order correlation function is a convenient tool to verify whether an electronic source emits a single-particle or a multi-particle state \cite{Moskalets:2015vr}. 
This function calculated for emitted particles is zero in the case of a single-particle state, while it differs from zero in the case of a multi-particle state. 

The total second-order correlation function, ${\cal G}^{(2)}\left(1,1^{\prime};2^{\prime},2 \right) = \langle \hat\Psi^{\dag}(1)\Psi^{\dag}(1^{\prime})  \hat\Psi(2^{\prime})\hat\Psi(2) \rangle$, is not additive. 
This means that it contains not only the Fermi sea contribution, ${\cal G}^{(2)}_{off}$, and the contribution of the particles emitted by the source, $G ^{(2)}$, but also the contributions due to the joint effect of emitted electrons and electrons of the Fermi sea.   
Therefore, the isolation of $G ^{(2)}$ is not a trivial problem anymore. 

When the state of emitted particles is pure this problem has a simple solution. 
Namely, as  suggested in Ref.~\cite{Moskalets:2014ea}, $G ^{(2)}$ is calculated as the Slater determinant of excess correlation functions $G^{(1)}$ taken at different time arguments: 

\begin{eqnarray}
\!\!\!\!\!\!\!\!\!\!\!\!
G^{(2)}\left(1,1^{\prime};2^{\prime},2  \right) = 
\det
\left ( 
\begin{array}{ll}
G^{(1)}(1,2)
&
G^{(1)}(1,2^{\prime})
\\
G^{(1)}(1^{\prime},2)
&
G^{(1)}(1^{\prime},2^{\prime})
\end{array}
 \right ) .
\label{34}
\end{eqnarray}
\ \\ \noindent
This suggestion works well at zero temperature \cite{Moskalets:2015vr}, whereas it fails at finite temperatures.
Indeed, if we use $G ^{(1)}_{V_{L}}$, Eq.~(\ref{28}), for a leviton emitted at a finite temperature, then Eq.~(\ref{34}) gives a non-zero result. 
This result disagrees with the expectation that $G ^{(2)}_{V_{L}} = 0$, since the state in question is a single-particle state.  

The resolution of this seeming paradox lies in the fact that with increasing temperature the state of a leviton evolves from pure to mixed. 
In the case of a pure state, Eq.~(\ref{34}) is correct. 
On the contrary, for a mixed state, $G ^{(2)}$ cannot be directly expressed in terms of $G ^{(1)}$ anymore. 
Different components of a mixed state should be considered separately while calculating a higher-order correlation function in terms of the Slater determinant composed of the first-order correlation functions. 

At finite temperatures, the correct expression for leviton's $G ^{(2)}_{V_{L}}$ is the following:

\begin{eqnarray}
\!\!\!\!\!\!\!\!\!\!\!\!
G^{(2)}_{V_{L}}\left(1,1^{\prime};2^{\prime},2  \right) = 
\int  d E \left( - \frac{   \partial f }{  \partial E } \right) G^{(2)}_{ V_{L},E}\left(1,1^{\prime};2^{\prime},2  \right) . 
\label{35} 
\end{eqnarray}
\ \\ \noindent
Here $G^{(2)}_{ V_{L},E}$ is expressed in terms of $G^{(1)}_{ V_{L},E}$, Eq.~(\ref{28}), according to Eq.~(\ref{34}). 
The equation above leads to $G^{(2)}_{V_{L}} = 0$ as expected for a single-particle  state. 

Note that in the general case, when an arbitrary periodic voltage $V(t)$ is applied to a metallic contact at a finite temperature, Eqs.~(\ref{28}) and (\ref{35}) define the first-order correlation function $G ^{(1)}_{V}$ and the second-order correlation function $G ^{(2)}_{V}$ for emitted particles respectively. 
In this case, the quantity $G ^{(1)}_{V,E}$ appearing in Eq.~(\ref{28}) reads:

\begin{eqnarray}
\label{eq:G1_ratio}
G ^{(1)}_{V,E}(t_{1}; t_{2}) =  \frac{e^{i \left( t_{1} - t_{2} \right) \frac{E }{ \hbar } } }{  v_{ \mu} } \frac{  e^{ i \frac{e }{ \hbar} \int _{ t_{2} }^{t_{1} } dt^{\prime} V(t^{\prime}) } - 1 }{2 \pi i \left(  t_{1} - t_{2}  \right) } .
\label{36}
\end{eqnarray}
\ \\ \noindent
In general, this quantity can not be factorized into the product of two terms dependent on one time only  as  done for $G^{(1)}_{ V_{L},E}( t_{1};t_{2})$. 
Therefore  $G ^{(2)}_{V} \ne 0$. This tells us that at finite temperatures a time-dependent voltage $V(t)$ excites a mixed multi-particle state on top of the Fermi sea.

\section{$G^{(1)}$ in presence of pure dephasing}
\label{sec5}

We take the opportunity of this work to emphasize differences in the coherence properties of single-electron states at finite temperatures or in presence of pure dephasing. In particular, investigating the purity and shot noise suppression in presence of pure dephasing will strengthen our previous interpretation of temperature-induced mixedness that it can not be considered equivalent to a decoherence effect. 

The model we consider here for pure dephasing is the following. 
\textit{At zero temperature}, we assume a finite region over which the single-electron state acquires an additional phase $\varphi(t)$ due to the presence of a classical potential $V(t)$. The time of flight for this region is defined as $t_V$ and the corresponding phase factor reads:

\begin{equation}
\label{eq:phase_factor}
e^{i \varphi(t)} =  e^{-i \frac{e }{\hbar } \int_{t-t_V}^t dt' V(t')}\,.
\end{equation}

\ \\ \noindent
The potential $V(t)$ is assumed to satisfy random Gaussian noise, \textit{i.e} 
\begin{equation}
\label{eq:def_V}
\langle V(t) \rangle = 0 \quad \text{and} \quad \langle V(t') V(t'') \rangle = 2 \left( \hbar/e \right)^{2}\delta(t-t')/\tau_\varphi\,.
\end{equation}
\ \\ \noindent
Here $\tau_\varphi$ is the dephasing time characterizing the fluctuations of the potential $V$. 

Our choice of modeling pure dephasing is in accordance with previous works in quantum optics investigating the effect of pure dephasing on single-photon sources 
\cite{Naesby08, Auffeves09, Cui06, Tichy15} 
and works investigating pure dephasing in mesoscopic physics, using a dephasing probe model introduced first by M. B\"uttiker \cite{Buttiker86,  Seelig01, Seelig03} or comparing different approaches for pure dephasing \cite{Brouwer97, Pilgram02, Marquardt_a, Marquardt_b, Chung05, Haack10}. This list of references is highly non exhaustive, it rather aims at highlighting the important contributions of M. B\"uttiker for a better understanding of quantum coherence and dephasing processes. 
Our model takes into account in an explicit way the fluctuations spectra of the environment, here the classical potential $V$. We have assumed a random Gaussian noise for simplicity, but the results presented below are generalizable to other spectra. 

The classical potential $V(t)$ affects the zero-temperature correlation function of a leviton, Eq.~(\ref{28}), in the following way,

\begin{equation}
\label{eq:def_g1}
G^{(1)}(t_1; t_2) = e^{-i \varphi(t_1)} e^{i \varphi(t_2)}   G^{(1)}_{ V_{L}, \mu}( t_{1};t_{2})\,.
\end{equation} 
\ \\ \noindent
If the time of flight is much longer compared to the leviton's lifetime, $t_{V} \gg  \Gamma _{\tau}$, then for all relevant times $t_{1}$ and $t_{2}$ the phase factor can be simplified as following:

\begin{eqnarray}
e^{-i (\varphi(t_1) - \varphi(t_2))} &=& e^{i \frac{e }{\hbar } (\int_{t_1-t_V} ^{t_1} dt' V(t') - \int_{t_2-t_V}^{t_2} dt' V(t'))} \nonumber \\ 
&\approx& e^{-i \frac{e }{\hbar } \int_{t_1}^{t_2} dt' V(t')}\,.
\end{eqnarray}
\ \\ \noindent
When averaging over the fluctuating phase $\varphi$, denoted $\langle \ldots \rangle_\varphi$, we assume Gaussian fluctuations such that:

\begin{eqnarray}
\langle e^{-i (\varphi(t_1) - \varphi(t_2)} \rangle_\varphi &=& e^{-\frac{1}{2} \langle (\varphi(t_1) - \varphi(t_2)^2  \rangle_\varphi} \nonumber \\
&\approx& e^{-\frac{1}{2}  \frac{e^{2} }{ \hbar^{2} } \iint_{t_1}^{t_2} dt' dt'' \langle V(t') V(t'') \rangle_\varphi} \nonumber \\
&=& e^{- \vert t_2 - t_1 \vert /\tau_\varphi}\,.
\end{eqnarray}
\ \\ \noindent
The last equality follows from the definition of the fluctuation potential $V$, see Eq.~(\ref{eq:def_V}). Inserting the expression of the scattering amplitude, Eq.~(\ref{27}), the final expression for the first-order correlation function of a leviton in presence of pure dephasing takes the form:

\begin{eqnarray}
\left\langle G^{(1)}(t_1, t_2) \right \rangle_\varphi = e^{-\vert t_2 - t_1 \vert/\tau_\varphi}  G^{(1)}_{ V_{L}, \mu}( t_{1};t_{2}) 
\nonumber \\
\label{eq:G1_deph} \\
= 
\int  d E \frac{  \hbar/( \pi \tau_{ \varphi}) }{ \left( E - \mu \right)^{2} + \left( \hbar/ \tau_{ \varphi} \right)^{2} }  G^{(1)}_{ V_{L},E}( t_{1};t_{2}) .
\nonumber 
\end{eqnarray}
\ \\ \noindent
As expected, the fluctuating phase factor affects the phase coherence properties of the quantum state, which can be probed via an interference current through the interferometer, Eq.~(\ref{12}).
This is reflected in the exponentially decaying factor in the first-order correlation function $G^{(1)}$. 
Let us remark that the measurement of $G^{(1)}$ should allow us to distinguish finite-temperature effect from pure dephasing, at least at sufficiently low temperatures, \textit{i.e.} when the thermal time $\tau_\theta$ exceeds the pulse duration $\Gamma_\tau$. Indeed, $G^{(1)}\left( t_{1}; t_{2} \right)$ will decay as a function of the time difference $\left | t_{1} - t_{2} \right |$ following the function $\eta\left( [t_{1} - t_{2}]/ \tau_{\theta} \right)$ introduced after Eq.~(\ref{31}),
whereas it will decay exponentially in presence of pure dephasing (see Eq.~(\ref{eq:G1_deph})). In the following, we make use of the possibility to measure $G^{(1)}$ in an experiment to investigate the purity of the single-electron state in presence of dephasing.

\subsection{Purity in presence of pure dephasing}

\begin{figure}[t]
\includegraphics[width=75mm]{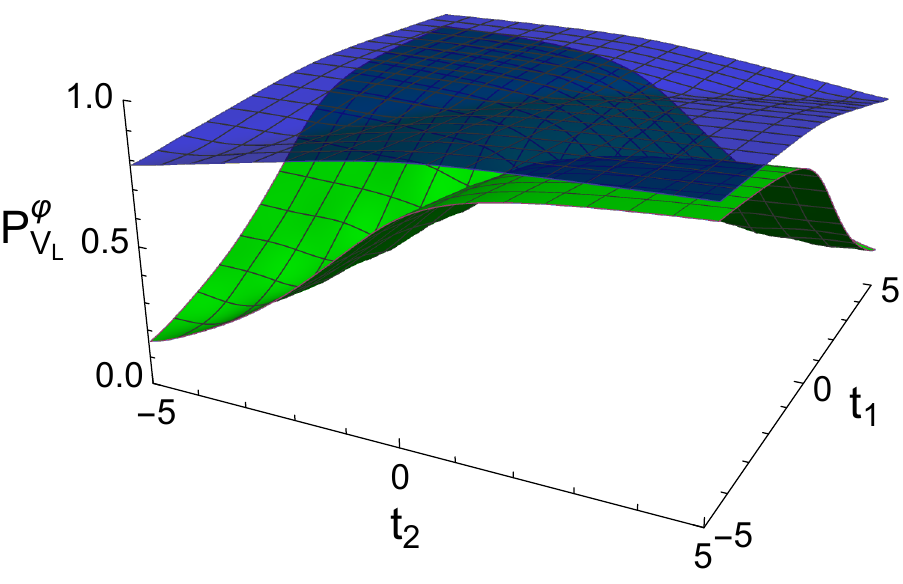}
\caption{(Color online) The purity coefficient $\mathrm{P_{V_{L}}^\varphi}$ for levitons emitted in presence of pure dephasing is shown as a function of  two times $t_{1}$ and $t_{2}$ normalized to the leviton's half-width $\Gamma_{ \tau}$, see Eq.~(\ref{p-deph}). The blue transparent surface calculated for a weak dephasing, $\tau_{ \varphi}=50\Gamma _{\tau}$, demonstrates that the state of a leviton is almost pure since the purity coefficient is close to one. While the green opaque surface calculated for a relatively strong dephasing, $\tau_{ \varphi}=5\Gamma _{\tau}$, demonstrates that the state is mixed since the purity coefficient deviates strongly from one. The maximum of leviton's density is at $t_{1} = t_{2} = 0$.}
\label{fig:purity_deph}
\end{figure}

We start again from the definition of the purity coefficient as defined in Eq.~(\ref{31}), with one important difference : as we are interested in the quantum state of a leviton subject to a fluctuating potential, the purity coefficient has to be calculated from the measured $G^{(1)}$ in presence of pure dephasing. This means that we have to consider $\left\langle G^{(1)}(t_1, t_2) \right \rangle_\varphi$ instead of $G^{(1)}(t_1, t_2)$. This leads to:

\begin{eqnarray}
\mathrm{P_{V_{L}}^\varphi} \left( t_{1}, t_{2} \right) &=& \frac{ v_{ \mu} \int dt  \left\langle  G ^{(1)}_{V_{L}}\left( t_{1}; t \right) \right \rangle_\varphi  \left\langle  G ^{(1)}_{V_{L}}\left( t; t_{2} \right) \right \rangle_\varphi }{ \left\langle  G ^{(1)}_{V_{L}}\left( t_{1}; t_{2} \right) \right \rangle_\varphi} \nonumber \\
\label{p-deph} \\
&=&  \int dt \frac{e^{-\vert t - t_1 \vert/\tau_\varphi}  e^{-\vert t_2 - t \vert/\tau_\varphi}}{e^{-\vert t_2 - t_1 \vert/\tau_\varphi} }  \frac{\Gamma_\tau/\pi  }{(t^2 +  \Gamma_\tau^2)} . \nonumber 
\end{eqnarray}
\ \\ \noindent
This purity coefficient is displayed  in Fig.~\ref{fig:purity_deph}: it shows that pure dephasing induces mixedness as expected for a finite dephasing time $\tau_\varphi$. 
Note that the effect of dephasing on a single-particle state is much stronger than the effect of finite temperature at close corresponding times $ \tau_{ \varphi}$ and $ \tau_{ \theta}$. 
This can be observed by comparing  the blue transparent surfaces in Fig.~\ref{purity} calculated for $\tau_{ \theta} = 10  \Gamma _{\tau}$ and in Fig.~\ref{fig:purity_deph} calculated for five times larger $ \tau_{ \varphi} = 50  \Gamma _{\tau}$.

Although the first-order correlation functions at finite temperatures and in presence of pure dephasing look similar, see respectively Eqs.~(\ref{28}) and (\ref{eq:G1_deph}), the difference in the underlying physical mechanisms which are responsible for mixedness results in completely different properties of the shot noise as we show below.

\subsection{Shot noise suppression in presence of pure dephasing}

Let us start with the situation where only one source emits single levitons as depicted on the left panel of Fig.Ã\ref{collider}. The excess current cross-correlation function averaged over the fluctuating phase $\varphi$ is then given by (see Eq.~(\ref{22})):

\begin{eqnarray}
\left \langle \frac{  {\cal P}^{ex} }{  {\cal P}_{0} } \right \rangle_\varphi 
&=& - v_{ \mu}^{2} \iint\limits dt dt^{\prime} \left \langle \left |   G ^{(1)}_{V_{L}}\left(t;t^{\prime} \right)   \right |^{2} \right\rangle_\varphi 
\nonumber \\
\\
&=& 
- v_{ \mu}^{2} \iint\limits dt dt^{\prime} \left |   G ^{(1)}_{V_{L} } \left(t;t^{\prime} \right)   \right |^{2} 
\nonumber 
\,.
\end{eqnarray}
\ \\ \noindent
The second equality follows from the cancellation of the phase factor $e^{-i(\varphi(t) - \varphi(t'))} e^{i (\varphi(t) - \varphi(t'))}$ for any $t$ and $t'$. Therefore, as we already mentioned, the shot noise is not affected by pure dephasing.

\begin{figure}[t]
\includegraphics[width=75mm]{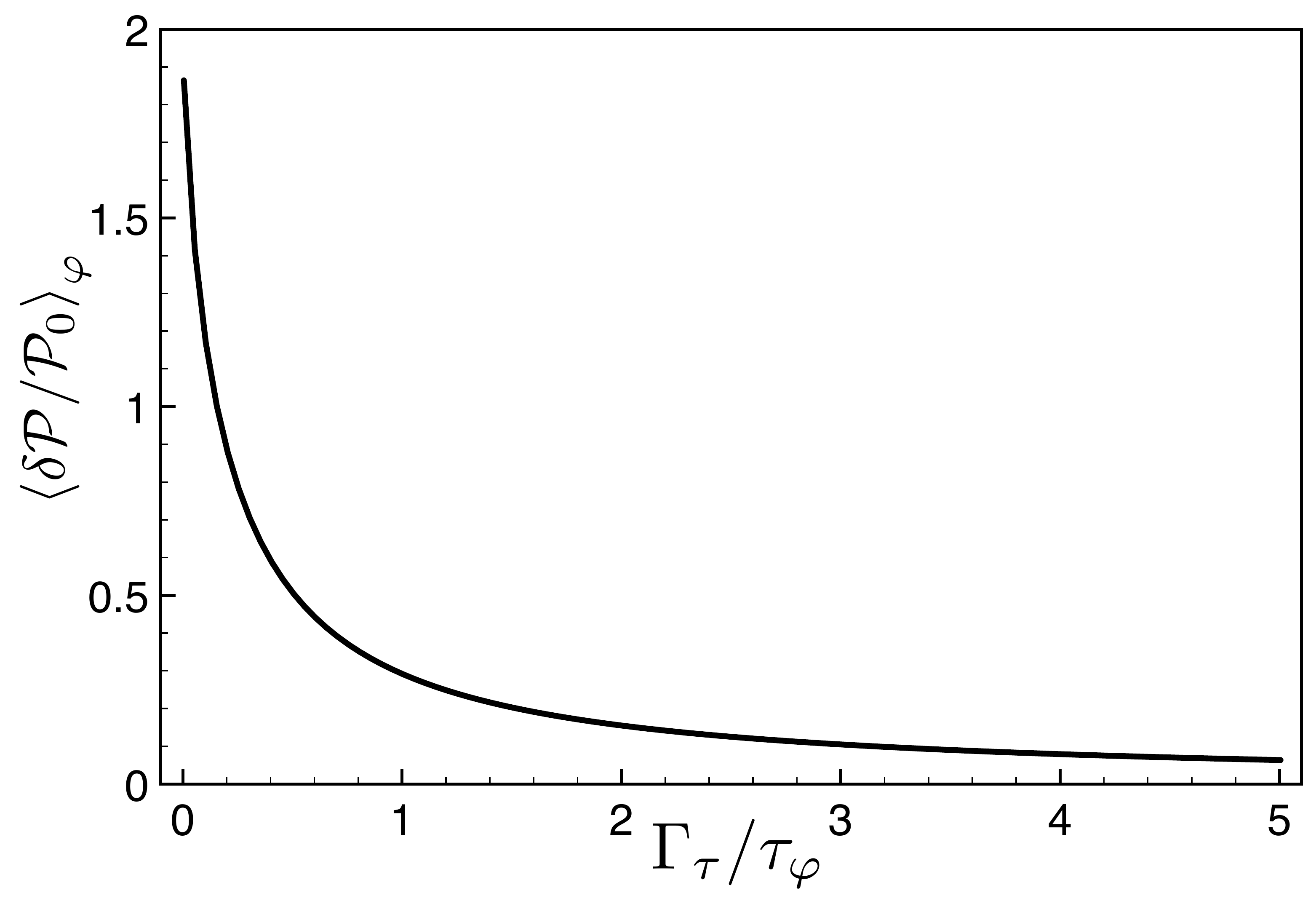}
\caption{The maximum correlation noise in presence of dephasing, $ \left\langle  \delta {\cal P} /  {\cal P}_{0}  \right\rangle_\varphi$,  see Eq.~(\ref{eq:noise2}), due to collisions of two identical levitons, $  G ^{(1)}_{1,V_{L}} =  G ^{(1)}_{2,V_{L}}$, is shown as a function of the ratio of the half-width of a leviton $  \Gamma _{\tau}$ and the dephasing time the same for both sources, $\tau_{ \varphi}\equiv \tau_{ 1,\varphi}=\tau_{ 2,\varphi}$.}
\label{fig:noise2_deph}
\end{figure}

The situation is radically different when considering shot noise arising from the antibunching of levitons emitted by two separated sources as shown on the right panel of Fig.~\ref{collider}. Indeed, in this situation, Eq.~(\ref{23}) becomes:

\begin{equation}
\left\langle \frac{  {\cal P}^{ex}_{two} }{  {\cal P}_{0} } \right\rangle_\varphi = - v_{ \mu}^{2} \iint\limits dt dt^{\prime} \left\langle \left\vert G^{(1)}_{1,V_{L} }(t,t') - G^{(1)}_{2,V_{L} }(t,t') \right \vert^2 \right\rangle_\varphi\,.
\end{equation}
\ \\ \noindent
The modulus squared of the difference between the two $G^{(1)}$-functions is composed of four terms. 
Assuming that the fluctuating potentials of two sources are not correlated and that they are characterized by the dephasing times $\tau_{1,\varphi}$ and $\tau_{2,\varphi}$, we finally obtain:

\begin{eqnarray}
\left\langle \frac{  {\cal P}^{ex}_{two} }{  {\cal P}_{0} } \right\rangle_\varphi &=& - 2  \left( 1 - \frac{1 }{2 } \left\langle  \frac{ \delta {\cal P} }{  {\cal P}_{0} } \right\rangle_\varphi \right) , 
\nonumber \\
\label{eq:noise2} \\
\left\langle   \frac{ \delta {\cal P} }{  {\cal P}_{0} } \right\rangle_\varphi
&=& 2 v_{ \mu}^{2} \iint\limits dt dt^{\prime} e^{-\vert t' - t \vert \left( \frac{1 }{ \tau_{1, \varphi} } + \frac{1 }{ \tau_{2, \varphi} } \right) } 
\nonumber \\
&& \times
G^{(1)}_{1,V_{L}  }(t,t') G^{(1)}_{2,V_{L} }(t',t)   \,.
\nonumber 
\end{eqnarray}
\ \\ \noindent
Equation~(\ref{eq:noise2}) shows that pure dephasing induces a suppression of the antibunching between two single-electron states at zero temperature 
in full agreement with Ref.~\cite{Iyoda:2014cf}. This result can also be understood in view of what was already mentioned earlier in this work: the correlation noise, $ \delta {\cal P}$ originates from a two-particle interference effect. When averaging over the fluctuating phase $\varphi$, the phase coherence is lost, which leads to the suppression of the correlation noise as shown in Fig.~\ref{fig:noise2_deph}.

This result has to be contrasted with the perfect antibunching of temperature-induced mixed states as presented in the previous section. The suppression of the correlation contribution to noise constitutes therefore a key measurement to distinguish a finite-temperature effect from pure dephasing.

\section{Conclusion}
\label{sec6}

We analyzed the first-order correlation function of electrons adiabatically injected on  top of the Fermi sea by a single-electron source. 
At finite temperatures of the Fermi sea, $\theta > 0$, the state of an injected  particle in mixed, while at zero temperature it is pure. 
The condition for the cross-over temperature is set by the thermal time $\tau_{ \theta} = \hbar / ( \pi k_{B} \theta )$ being of the order of the lifetime of a single-particle wave-packet $2  \Gamma _{\tau}$, see Fig.~\ref{purity}.
We showed that despite its mixed nature, the single-particle states remain phase coherent at finite temperatures. 
This is demonstrated by the fact that two such states (with identical characteristics) show perfect antibunching while overlapping at an electronic wave splitter. 
As a result of antibunching, the shot noise vanishes if the two states approach the wave splitter from different sides at the same time.  
This effect is an electronic analogue of the famous Hong-Ou-Mandel effect, where bunching of two photons nullifies the correlation function for outgoing streams of particles. 
Importantly, here we demonstrated that even fermions in a temperature-induced mixed states demonstrate perfect antibunching. 
This result was emphasized by considering single-particle states at zero temperature but subject to pure dephasing. In this case, their phase coherence properties are altered and we have shown that this leads to a suppression of the antibunching.

However the mixed nature of a state injected at finite temperature has an  observable effect. 
In particular, a mixed single-particle state produces less shot noise caused by partitioning at the wave splitter, compared to what is produced by a pure single-particle state. As a result, the shot noise gets suppressed with increasing temperature in agreement with the experimental observation of Ref.~\cite{Bocquillon:2012if}.

\section*{Acknowledgments}

We dedicate this work to Markus B\"uttiker, who has always been a great source of inspiration through his enthusiasm and critical questions. We thank Christian Glattli and Mathias Albert for useful comments on the manuscript. M.M. acknowledge the financial support from and the warm hospitality of the \'Ecole Normale Sup\'erieure de Lyon where part of this work was done. G.H. acknowledges support from the ERC grant MesoQMC.



\end{document}